\documentstyle[aps,prl,preprint]{revtex}
\draft

\begin{document}
\author{Wei-Min Sun, Xiang-Song Chen and Fan Wang}
\address{Department of Physics and Center for Theoretical Physics, Nanjing University, Nanjing 210093, China}
\title{A Note on Functional Integral over the Local Gauge Group}
\maketitle

\begin{abstract}
We evaluated some particular type of functional integral over the local gauge group 
$C^{\infty}({\bf R}^n, U(1))$ by going to a discretized lattice. The results explicitly violates the property of the
Haar measure. We also analysed the Faddeev-Popov method through a toy example.
The results also violates the property of the Haar measure.
\end{abstract}

Functional integral is a powerful tool in quantum theory, especially QFT. 
In this paper we address the problem of functional integral over the local gauge
group. We choose to study this topic because functional integral over the gauge
group plays a decisive role in the Faddeev-Popov method \cite{Faddeev}of quantizing
 a gauge field
theory. On the one hand in the Faddeev-Popov method the existence of a Haar measure
  on the
local gauge group is implicitely assumed. On the other hand the local gauge group is
infinite dimensional and not locally compact, and hence does not necessarily posess
a Haar measure. As in the case of ordinary functional integral, we could take
functional integral over the local gauge group as some limit of finite dimensional
integrals evaluated on a discretized lattice. In a discretized lattice a Haar
measure obviously exists. Then we also have to study whether such a limit exists or
not. 

In this paper we evaluated a particular type of functional integration on the local
gauge group $C^{\infty}({\bf R}^n,U(1))$ by going to a discretized lattice. The
results are not in favor of a Haar measure, as will be shown in the following. 

Let $\omega$ stand for elements of $C^{\infty}({\bf R}^n, U(1))$. $f(z)$ is an
entire function which is bounded on the imaginary axis $\{z \in {\bf C}| Re(z)=0\}$.
We consider the following continuous functional of $\omega$: 
\begin{equation}
F[\omega]=f(\omega^{-1}(x)\partial_{\mu}\omega(x))
\end{equation}
Here $x$ is a fixed point in ${\bf R}^n$. If we write $\omega(x)=e^{i\theta(x)}$ we
can see that $\omega^{-1}(x)\partial_{\mu}\omega(x)=i\partial_{\mu}\theta(x)$ lies
on the imaginary axis and so that $F[\omega]$ is bounded.. If a Haar measure exists the volume of $C^{\infty}({\bf R}^4,U(1))$ could be taken to
be $1$ and $F[\omega]$ is integrable with respect to it. Now
let us evaluate $\int_G D\omega F[\omega]$ by going to a discretized lattice. Obviously
we should make the replacement: $\partial_{\mu}\omega(x) \rightarrow
\frac{\omega(x+\epsilon)-\omega(x)}{\epsilon^{\mu}}~~or~~\frac{\omega(x+\epsilon)-
\omega(x-\epsilon)}{2\epsilon^{\mu}}$. If we make the first replacement we can write
\begin{eqnarray}
\int_G D\omega F[\omega] &=& \int_G D\omega f(\omega^{-1}(x)\partial_{\mu}\omega(x))
 \nonumber \\
&=& \lim_{\epsilon^{\mu} \rightarrow 0}\int \prod_{i}d \omega_{i}
f(\frac{\omega^{-1}(x)\omega(x+\epsilon)-1}{\epsilon^{\mu}}) \nonumber \\
&=& \lim_{\epsilon^{\mu} \rightarrow 0}\int \prod_{i}d \omega_{i}
\sum_{n=0}^{\infty}\frac{1}{n!}f^{(n)}(0)(\frac{\omega^{-1}(x)\omega(x+\epsilon)-1}
{\epsilon^{\mu}})^n \nonumber \\
&=& \lim_{\epsilon^{\mu} \rightarrow 0}\sum_{n=0}^{\infty}\frac{1}{n!}f^{(n)}(0)\int
\prod_{i}d \omega_{i}(\frac{\omega^{-1}(x)\omega(x+\epsilon)-1}{\epsilon^{\mu}})^n 
\nonumber \\
&=& \lim_{\epsilon^{\mu} \rightarrow 0}\sum_{n=0}^{\infty}\frac{1}{n!}f^{(n)}(0)(-
\frac{1}{\epsilon^{\mu}})^n \nonumber \\
&=& \lim_{\epsilon^{\mu} \rightarrow 0}f(-\frac{1}{\epsilon^{\mu}}) \nonumber \\
&=& f(\infty)
\end{eqnarray}
In the fourth line we have interchanged the order of integration and summation. This
is legitimate because $|\frac{\omega^{-1}(x)\omega(x+\epsilon)-1}{\epsilon^{\mu}}|
\leq \frac{2}{\epsilon^{\mu}}$ and the Taylor series of $f(z)$ is uniformly
convergent in any closed disc $|z| \leq R$. In obtaining the result of the fifth
line we have used the formulas $\int_{U(1)}d\omega \omega^n = \delta_{n0}$,which can
be easily verified.
If we note that
\begin{eqnarray}
F[\omega_0\omega]&=& f((\omega_0(x)\omega(x))^{-1}\partial_{\mu}(\omega_0(x)\omega(x))) \nonumber \\
&=& f(\omega_0^{-1}(x)\partial_{\mu}\omega_0(x)+\omega^{-1}(x)\partial_{\mu}\omega(x))
\end{eqnarray}
we can immediately obtain
\begin{eqnarray}
\int_G D\omega F[\omega_0\omega]&=& f(\omega_0^{-1}(x)\partial_{\mu}\omega_0(x)+\infty) \nonumber \\
&=& f(\infty+i\partial_{\mu}\theta_0(x))
\end{eqnarray}
We can see that the property of the Haar measure could be violated.

If we use the second replacement for $\partial_{\mu}\omega(x)$ in evaluating $\int_G
D\omega F[\omega]$ and $\int_G D\omega F[\omega_0\omega]$  we will see that there is
also an obvious violation of the property of the Haar measure. To see this the only thing we
 should change is to
replace $\int
 \prod_{i}d\omega_{i}(\frac{\omega^{-1}(x)\omega(x+\epsilon)-1}{\epsilon^{\mu}})^n$ 
by $\int \prod_{i}d\omega_{i}(\frac{\omega^{-1}(x)\omega(x+\epsilon)-\omega^{-1}(x)
\omega(x-\epsilon)}{2\epsilon^{\mu}})^n$ in Eq(2). The latter integral is $\delta_
{n0}$. So the integral $\int_G D\omega F[\omega]$ is evaluated to be $f(0)$.
Similarly the integral $\int_G D\omega F[\omega_0\omega]$ equals $f(i\partial_{\mu}
\theta_0(x))$.

Now we suppose $f(z)=e^{a^2 z^2},a \in {\bf R}-\{0\}$. This function satisfies the
conditions we imposed earlier. Then in the first discretization scheme $\int_G
D\omega F[\omega]$ is $\infty$ while $\int_G D\omega f[\omega_0\omega]$ has an
oscillating phase and an indefinitely increasing modulus. The integration value for
$F[\omega_0\omega]$ is unreasonable because we have a real-valued integrand. In
the second discretization scheme both integration value are real and finite but not
equal, thus
violating the property of the Haar measure.

Then what is the origin of the incorrectness of the first discretization scheme?
Let us expand $\omega^{-1}(x)\frac{\omega(x+\epsilon)-\omega(x)}{\epsilon}$ and
$\omega^{-1}(x)\frac{\omega(x+\epsilon)-\omega(x-\epsilon)}{2\epsilon}$ in $\epsilon
$(here we have suppressed the superscript $\mu$ for clarity). 
\begin{eqnarray}
\omega^{-1}(x)\frac{\omega(x+\epsilon)-\omega(x)}{\epsilon} &=& \omega^{-1}(x)\omega
'(x)+\frac{1}{2}\omega^{-1}(x)\omega''(x)\epsilon+O(\epsilon^2) \nonumber \\
\omega^{-1}(x)\frac{\omega(x+\epsilon)-\omega(x-\epsilon)}{2\epsilon} &=&
\omega^{-1}(x)\omega'(x)+O(\epsilon^2)
\end{eqnarray}
Both finite differences tends to the purely imaginary number
$\omega^{-1}(x)\omega'(x)$ in the limit $\epsilon \rightarrow 0$, i.e.,they have
the same continuum limit. But in the first discretization scheme the speed of the
finite difference
tending to its continuum limit is first order and the coefficient $\frac{1}{2}
\omega^{-1}(x)\omega''(x)$ does not lie on the imaginary axis, thus leading to the
unreasonble results for $\int_G D\omega f[\omega_0\omega]$. In the second
discretization scheme the speed of the finite difference tending to the continuum
limit is second order hence has not lead to unreasonable results for $\int_G
D\omega F[\omega]$ and $\int_G D\omega F[\omega_0\omega]$. So the first
discretization scheme is unsuitable and should be abandoned. In the following
we will take the second discretization scheme.
   
Up to now we have studied the functional integral of some particular type of
functionals of $\omega$. In the Faddeev-Popov method we meet with a
$\delta$-functional. With the above obtained results in hand we can also study
whether similar problems arise in the F-P method.

In the F-P method the following equility is used to define the gauge-invariant 
Faddeev-Popov determinant $\Delta_G[A]$.
\begin{equation}
\Delta_G[A]\int_G D\omega \delta(G(A^{\omega}))=1
\end{equation}
Here $G(A;x)$ is a gauge fixing function. In the Lorentz gauge which we shall
take in the following, we have
$G(A;x)=\partial_{\mu}A^{\mu}(x)$. $A^{\omega}$ is the gauge transformation of $A$.
Written out expicitly, $A_{\mu}^{\omega}(x)=A_{\mu}(x)-\partial_{\mu}\theta(x)$.
The $\delta$-functional appearing in Eq(6) is understood formally as $\prod_{x} 
\delta(\partial_{\mu}A^{\mu}(x)-\Box \theta(x))$. Rigorously a $\delta$-function
should be define as some (continuous) linear form on some function space. We will adopt this definition in the following discussion. For technical simplicity 
we will assume $n=1$, i.e., the gauge group is $C^{\infty}({\bf R},U(1))$. We also assume $\lim_{x
 \rightarrow
\infty}\omega(x)=1$. This will not cause any essential difference. 
The gauge field $A_{\mu}(x)$ now has only one component $A(x)$(of course this 
does not correspond to any physical reality; but here we use it just as a toy example).
We will assume $A(x)$
tends to $0$ at spatial infinity. Now the $\delta$-functional is defined to be a
linear form on the space of fast decreasing test functionals of $A$. Written out explicitly: $\langle 
\delta, T\rangle=T[0]$. The gauge fixing condition $G(A^{\omega})=0$ reads:
\begin{equation}
A'(x)-\theta''(x)=0
\end{equation}
If we assume $A(x)$ tends to $0$ at spatial infinity the above equation has only one
solution $A(x)=\theta'(x)$.(we have assumed that $\lim_{x \rightarrow
\infty}\omega(x)=1$ and this implies $\lim_{x \rightarrow \infty}\theta'(x)=0$.)

Obviously the $\delta$-functional in Eq(6) should be the following linear form:
$\langle \delta_{\omega},T\rangle=T[\theta']$. If we take $T[A]$ to be 
the functional $e^{-a^2 A^2(x)}$ which is fast decreasing for large $A$, we have
\begin{eqnarray}
\langle \int_G D\omega \delta_{\omega},T\rangle &=& \int_G D\omega\langle
\delta_{\omega},T\rangle \nonumber \\
&=& \int_G D\omega T[\theta']
\end{eqnarray}
If we take $f(z)=e^{a^2z^2}$ as above we see immediately that $T[\theta']$  is
just $F[\omega]$ defined earlier. In virtue of the above obtained results we
conclude that $\int_G D\omega \delta_{\omega} \not= \int_G D\omega
\delta_{\omega_0\omega}$.

From the above calculation we can obviously see that there must be some problems.
As we have mentioned earlier, the local gauge group does not necessarily posess
a Haar measure. If a Haar measure really does not exist, then the above situation 
is no surprise. But the Faddeev-Popov method will be in danger. Physicists usually
assume a Haar measure exists on the local
gauge group. Then we should say that the usual discretization method is in danger.
This is not the case in the usual path integral fourmalism of quantum mechanics, where a
discretization procedure does produce the correct quantum mechanical transition
amplitude \cite{Itzykson}.
  
Our conclusion is that on the local gauge group either a Haar measure does not
exist, or the usual discretization method of defining functional integral should be
modified(we are inclined to think that the former possibility is more probable
to occur).
 This of course is due to the complexity of
measures in infinite dimensional spaces.

We thank Feng-Wen An for useful discussions. This work is supported by NSF, SED and SSTC of China.

\end{document}